\documentclass[12pt]{article}
\usepackage{amsfonts}
\usepackage{amsmath}
\usepackage{amssymb}

\numberwithin{equation}{section} \numberwithin{theorem}{section}
\makeatother

\begin{document}

\title{From Random Matrices to Quasiperiodic Jacobi Matrices via Orthogonal
Polynomials}
\author{L. Pastur \\
Institute for Low \ Temperature Physics, Kharkiv, Ukraine}
\date{}
\maketitle


\begin{abstract}
\noindent We present an informal review of results on asymptotics
of orthogonal polynomials, stressing their spectral aspects and
similarity in two cases considered. They are polynomials
orthonormal on a finite union of disjoint
intervals with respect to the Szeg\"o weight and polynomials orthonormal on $%
\mathbb{R}$ with respect to varying weights and having the same union of
intervals as the set of oscillations of asymptotics. In both cases we
construct double infinite Jacobi matrices with generically quasiperiodic
coefficients and show that each of them is an isospectral deformation of
another. Related results on asymptotic eigenvalue distribution of a class of
random matrices of large size are also shortly discussed.
\end{abstract}

\section{Introduction}
\label{intr}

The goal of the paper is to discuss a link between asymptotics of a
class of orthogonal polynomials, in particular polynomials with
respect to varying weights (see e.g. \cite{To:94}), and the Jacobi
matrices with quasi-periodic coefficients, seen mostly as a
particular case of ergodic finite-difference operators. The theory
of this class of operators owes a lot to B. Simon, starting from an
early book \cite{Cy-Co:87} till just appeared impressive
\cite{Si:05}. The link became clear while the author was reflecting
on applications of the asymptotic formulas, found in the remarkable
paper by P. Deift et al \cite{De-Co:99}, to certain problems on the
eigenvalue distribution of a class of random matrices, known as
unitary invariant matrix models. This is why we would like to begin
from a discussion of random matrices, despite that the link can be
described without a recourse to random matrices.

Consider $n\times n$ Hermitian random matrices
\begin{equation}
M_n=\{M_{jk}\in \mathbb{C},\;M_{kj}=\overline{M_{jk}}\}_{j,k=1}^{n},
\label{M}
\end{equation}%
whose probability law is
\begin{equation}
P(dM_n)=Z_{n}^{-1}\exp \{-n\mathrm{Tr}V(M_n)\}dM_n.  \label{MM}
\end{equation}%
Here $Z_{n}$ is the normalization constant, $V:\mathbb{R}\rightarrow \mathbb{%
R}_{+}$ is a continuous, bounded below and growing at infinity function
(think about a polynomial of an even degree, positive at infinity, see also (%
\ref{Vlog})), and
\begin{equation}
dM_n=\prod\limits_{j=1}^{n}dM_{jj}\prod\limits_{1\leq j<k\leq
n}^{n}d\Re M_{jk}d\Im M_{jk}.  \label{dM}
\end{equation}%
This class of random matrices arises in a number of fields of mathematics
and physics (see e.g. reviews \cite%
{DiF:95,We-Co:98,Ka-Sa:99,Me:92,Pa:96,Pa:00} and references
therein). A considerable amount of corresponding problems can be
described in terms of the Normalized Counting Measure of eigenvalues
(NCM), defined as the relative to $n$ number of eigenvalues of
$M_n$, falling into a given set
$\Delta \subset \mathbb{R}$:%
\begin{equation}
N_{n}(\Delta )= \sharp \{\lambda _{l}^{(n)}\in \Delta
,\;l=1,...,n\}\big/ n,  \label{NCM}
\end{equation}%
where%
\begin{equation}
\{\lambda _{l}^{(n)}\}_{l=1}^{n}  \label{ev}
\end{equation}%
are eigenvalues of $M_n$.

We note that similar measures arise in spectral theory of ergodic
operators, and B. Simon did a lot of excellent work on the
measures, whose limit as $n\rightarrow \infty $ is known there as
the Integrated Density of States.

It will be convenient to consider a bit more general object, known as a
linear eigenvalue statistics and defined via a test function $%
\varphi :\mathbb{R}\rightarrow \mathbb{R}$ and eigenvalues (\ref{ev}):
\begin{equation}
N_{n}[\varphi ]=n^{-1}\sum_{l=1}^{n}\varphi (\lambda _{l}^{(n)}).
\label{stl}
\end{equation}%
We obtain (\ref{NCM}) by setting $\varphi =\chi _{\Delta }$, where
$\chi _{\Delta }$ is the indicator of $\Delta $.

Here are three basic quantities of the Random Matrix Theory, related to
eigenvalue statistics and widely studied, especially as $n\rightarrow \infty
$.
\begin{enumerate}
\item[(i)] Expectation of $N_{n}[\varphi ]$ with respect to (\ref{MM}) -- (\ref%
{dM}):%
\begin{equation}
\overline{N}_{n}[\varphi ]=\mathbf{E}\{N_{n}[\varphi ]\}.
\label{Nbar}
\end{equation}
\item[(ii)] Covariance of $N_{n}$ for two test functions $\varphi _{1.2}$:%
\begin{eqnarray}\label{Cov}
&&\mathbf{Cov}\{N_{n}[\varphi _{1}],N_{n}[\varphi _{2}]\}\\&&
\hspace{1cm}=\mathbf{E}%
\{N_{n}[\varphi _{1}]N_{n}[\varphi _{2}]\}-\mathbf{E}\{N_{n}[\varphi _{1}]\}%
\mathbf{E}\{N_{n}[\varphi _{2}]\}.  \nonumber
\end{eqnarray}
\item[(iii)] Gap probability%
\begin{equation}
E_{n}(\Delta )=\mathbf{P}\{N_{n}(\Delta )=0\}.  \label{ggap}
\end{equation}
\end{enumerate}
\noindent It is a remarkable observation by Gaudin, Mehta, and
Dyson of the early 60th (see e.g. \cite{Me:92}) that the above
quantities can be expressed via the orthonormal polynomials
$\{p_{l}^{(n)}\}_{l\geq 0}$ with respect to the
weight%
\begin{equation}
w_{n}(\lambda )=e^{-nV(\lambda )},  \label{wn}
\end{equation}%
\begin{equation}
\int w_{n}(\lambda )p_{l}^{(n)}p_{m}^{(n)}d\lambda =\delta
_{lm},\;l,m=0,1,...  \label{ortho}
\end{equation}%
Here and below integrals without limits denote integrals over $\mathbb{R}$.
We will call $V$ potential. Denote%
\begin{equation}
\psi _{l}^{(n)}(\lambda )=w_{n}^{1/2}(\lambda )p_{l}^{(n)}(\lambda ),
\label{psi}
\end{equation}%
and
\begin{equation}
K_{n}(\lambda ,\mu )=\sum_{l=0}^{n-1}\psi _{l}^{(n)}(\lambda )\psi
_{l}^{(n)}(\mu ).  \label{Kn}
\end{equation}%
$K_{n}$ is called the reproducing kernel of the orthonormal system $\{\psi
_{l}^{(n)}(\lambda )\}_{l\geq 0}$, and $(K_{n}(\lambda ,\lambda ))^{-1}$ is
known in the approximation theory as the Christoffel function \cite%
{Ak:62,Sa-To:97,To:94}.

To distinguish these polynomials from the traditional ones for which the
weight does not contain the large parameter $n$, the polynomials (\ref{wn})
-- (\ref{psi}) are called the orthogonal polynomials with respect to varying
weights (see e.g. \cite{To:94}).

We have for (\ref{Nbar}) -- (\ref{ggap}):%
\begin{equation}
\overline{N}_{n}[\varphi ]=\int \varphi (\lambda )\rho _{n}(\lambda
)d\lambda ,\;\rho _{n}(\lambda )=n^{-1}K_{n}(\lambda ,\lambda ),
\label{NbK}
\end{equation}%
see \cite{Me:92},
\begin{equation}
\mathbf{Cov}\{N_{n}[\varphi _{1}],N_{n}[\varphi _{2}]\}=\frac{1}{2n^{2}}\int
\int \frac{\Delta \varphi _{1}}{\Delta \lambda }\frac{\Delta \varphi _{2}}{%
\Delta \lambda }K_{n}^{2}(\lambda _{1},\lambda _{2})d\lambda _{1}d\lambda
_{2},  \label{CovK}
\end{equation}%
where%
\begin{equation}
\frac{\Delta \varphi }{\Delta \lambda }=\frac{\varphi (\lambda
_{1})-\varphi (\lambda _{2})}{\lambda _{1}-\lambda _{2}},
\label{dede}
\end{equation}%
see \cite{Pa-Sh:97}, and
\begin{equation}
E_{n}(\Delta )=\det (1-K_{n}(\Delta )),  \label{detK}
\end{equation}%
where $K_{n}(\Delta )$ is the integral operator%
\begin{equation}
(K_{n}(\Delta )f)(\lambda )=\int_{\Delta }K_{n}(\lambda ,\mu )f(\mu
)d\mu ,\;\lambda \in \Delta,  \label{Kker}
\end{equation}%
see \cite{Me:92}.

We will need two more basic facts on orthonormal polynomials. The
first is the Jacobi matrix $J^{(n)}$, associated with the
polynomials via
the r.h.s. of the three term recurrence relation:%
\begin{equation}
\lambda p_{l}^{(n)}(\lambda )=r_{l}^{(n)}p_{l+1}^{(n)}(\lambda
)+s_{l}^{(n)}p_{l}^{(n)}(\lambda
)+r_{l-1}^{(n)}p_{l-1}^{(n)}(\lambda ), \; r_{-1}=0, \label{ur}
\end{equation}%
i.e.,
\begin{equation}
J^{(n)}=\{J_{lm}^{(n)}\}_{l,m=0}^\infty, \quad
J_{lm}^{(n)}=r_{l}^{(n)}\delta _{l+1,m}+s_{l}^{(n)}\delta
_{l,m}+r_{l-1}^{(n)}\delta _{l-1,m}. \label{Jn}
\end{equation}%
The second is the Christoffel-Darboux formula%
\begin{equation}
K_{n}(\lambda ,\mu )=r_{n-1}^{(n)}\frac{\psi _{n}^{(n)}(\lambda )\psi
_{n-1}^{(n)}(\mu )-\psi _{n-1}^{(n)}(\lambda )\psi _{n}^{(n)}(\mu )}{\lambda
-\mu }.  \label{CDf}
\end{equation}%
Formulas (\ref{NbK}) -- (\ref{Kker}) and (\ref{CDf}) show that the
asymptotic form of (\ref{Nbar}) -- (\ref{ggap}) as $n\rightarrow \infty $ is
determined by that of $\psi _{n-1}^{(n)}$ and $\psi _{n}^{(n)}$. These were
found by Deift et al \cite{De-Co:99} in the case, where $V$ is real analytic
(see also \cite{Bl-It:97,Da-Co:00}). In the same paper the limit of (\ref%
{detK}) was found in the so-called local asymptotic regime (see e.g. \cite%
{Pa:00} for its definition, and \cite{Pa-Sh:97} for another
derivation of this result).

The author of this paper has applied the results of \cite{De-Co:99} to find
an asymptotic form of the covariance (\ref{Cov}), and to prove an analog of
the central limit theorem for linear eigenvalue statistics (\ref{stl}). It
turned out that this requires a bit more information on the asymptotic
formulas of \cite{De-Co:99}, and leads to certain objects, related to
spectral theory of quasiperiodic Jacobi matrices. This is discussed in the
paper. Asymptotics of covariance and an analog of the central limit theorem
for linear statistics of eigenvalues of random matrices (\ref{MM}) -- (\ref%
{dM}) will be published elsewhere \cite{Pa:05}.

The paper is organized as follows. In the next section asymptotic
formulas for the "ordinary" polynomials orthogonal with respect to
weights whose support is a union of $q\geq 1$ disjoint intervals
are shortly discussed following papers
\cite{Ak:60,Ap:84,Pe-Yu:03,Wd:69}. We present then asymptotics
found in \cite{De-Co:99} for polynomials, orthogonal with respect
to varying weights for the case, where their oscillatory part is
the same union of $q$ intervals. In Section 3 we introduce
quasiperiodic Jacobi matrices, associated with the both
asymptotics and discuss links between the matrices. We argue that
they are so-called "finite-band" Jacobi matrices, widely known in
the theory of integrable systems, that they are related by an
isospectral deformation, and consider a particular case of
polynomial potentials in (\ref{wn}), where corresponding Jacobi
matrices are periodic. In Section 4 we present a collection of
facts on asymptotic eigenvalue distributions of random matrices,
that can be written in the terms of the above Jacobi matrices. In
Appendix we give a direct proof of the isospectrality of the
Jacobi matrices related to asymptotics of both classes of
orthogonal polynomials.

\section{Asymptotics of orthogonal polynomials}

\subsection{Ordinary orthogonal polynomials}

Consider first the case, where the weight $w$ does not depend on $n$. In
this case all the quantities, related to orthonormal polynomials, do not
depend on the super-index $(n)$. We will denote the polynomials and related
quantities by the same symbols as in (\ref{ortho}) -- (\ref{Kn}) and (\ref%
{ur}) -- (\ref{CDf}) but without the super-index $(n)$. Assume that the
support $\sigma $ of the weight is a finite union of disjoint finite
intervals:%
\begin{equation}
\sigma =\bigcup_{l=1}^{q}[a_{l},b_{l}],\;-\infty
<a_{1}<b_{1}<...<a_{q}<b_{q}<\infty.  \label{sigs}
\end{equation}%
Denote $\mathcal{M}_{1}(\sigma)$ the set of non-negative unit
measures
on $\sigma$ and consider the quadratic functional%
\begin{equation}
\mathcal{E}_{\sigma }[m]=-\int_{\sigma \times \sigma }\log |\lambda
-\mu |m(d\lambda )m(d\mu ),\;m\in \mathcal{M}_{1}(\sigma).
\label{Es}
\end{equation}%
The functional possesses a unique minimizer $\nu$
(the equilibrium measure for $%
\sigma $):%
\[
\min_{m\in \mathcal{M}_{1}(\sigma)}\mathcal{E}_{\sigma }[m]=\mathcal{E}%
_{\sigma }[\nu ].
\]%
This is a standard variational problem of
potential theory, that admits a simple electrostatic
interpretation in which $m$ is a distribution of positive charges
on a conductor $\sigma$ and $\nu$ is the equilibrium distribution
of charges.

It is known (see e.g. \cite{Sa-To:97}) that the problem is
equivalent to the relations%
\begin{equation}
-2\int_{\sigma }\log |\lambda -\mu |\nu (d\mu )=-l_{\sigma },\;\lambda \in
\sigma ,  \label{Es1}
\end{equation}%
\begin{equation}
-2\int_{\sigma }\log |\lambda -\mu |\nu (d\mu )\geq -l_{\sigma
},\;\lambda \in \mathbb{R}\setminus \sigma ,  \label{Es2}
\end{equation}%
that are the Euler-Lagrange equations for (\ref{Es}). The quantity $%
-l_{\sigma }/2$ is known as the Robin constant, and $e^{l_{\sigma }/2}$ is
the logarithmic capacity of $\sigma $.

Set%
\begin{equation}
\nu (\lambda )=\nu ((\lambda ,\infty )),  \label{nu}
\end{equation}%
and
\begin{equation}
\alpha =\{\alpha _{l}\}_{l=1}^{q-1},\;\alpha _{l}=\nu (a_{l+1}).
\label{alpha}
\end{equation}%
With this notation the asymptotic formulas of \cite{Ap:84},
Theorem 1, (see also \cite{Ak:60,AT:61,Pe-Yu:03,Wd:69}) for
analogs of orthonormalized functions (\ref{psi}), corresponding to
a $n$-independent weights $w$, can be written as follows.

Assume that the weight satisfies the Szeg\"{o} condition%
\begin{equation}
\log w\in L^{1}(\sigma, \nu ).  \label{Sz}
\end{equation}%
Then there exist the functions $\mathcal{D}_{\sigma
}:\sigma \times \mathbb{T}^{q-1}\rightarrow \mathbb{R}_{+}$, and $\;\mathcal{%
G}_{\sigma }:\sigma \times \mathbb{T}^{q-1}\rightarrow \mathbb{R}$ such that
if $\lambda $ belongs to the interior of $\sigma $, we have
\begin{eqnarray}\label{psis}
\psi _{n}(\lambda )&=&(2\mathcal{D}_{\sigma }(\lambda ,n\alpha ))^{1/2}
\\ &\times& \cos %
\Big(\pi n\nu (\lambda )+\mathcal{G}_{\sigma }(\lambda ,n\alpha )\Big)%
+o(1),\quad n\rightarrow \infty , \nonumber
\end{eqnarray}%
where the remainder vanishes in the $L^{2}(\sigma )$-norm, and
\begin{equation}
n\alpha =(n\alpha _{1},...n\alpha _{q-1}).  \label{dgan}
\end{equation}%
Besides, there exist functions $\mathcal{R}_{\sigma }:\mathbb{T}%
^{q-1}\rightarrow \mathbb{R}_{+}$, and $\;\mathcal{S}_{\sigma }:\mathbb{T}%
^{q-1}\rightarrow \mathbb{R},$ such that the coefficients $\{r_l,
s_l \}_{l\ge 0}$ of the corresponding Jacobi matrix $J_{\sigma }$
(\ref{Jn}), that does not depend on the super-index $n$ in this
case, have the following asymptotic form
\begin{equation}
r_{n}=\mathcal{R}_{\sigma }(n\alpha )+o(1),\;s_{n}=\mathcal{S}_{\sigma
}(n\alpha )+o(1),\;n\rightarrow \infty ,  \label{rsRSs}
\end{equation}%
Note that to find (\ref{rsRSs}) one needs weaker asymptotics of $%
p_{n}(\lambda )$, those for $\lambda $ outside $\sigma $.


Functions $\mathcal{D}_{\sigma },%
\mathcal{G}_{\sigma },\mathcal{R}_{\sigma }$, and
$\mathcal{S}_{\sigma }$ can be expressed via the $(q-1)$ -
dimensional Riemann theta-function (see e.g. formula (\ref{Rx})
below), associated with the two-sheeted Riemann surface. The
surface is obtained by gluing together two copies of the complex
plane slit along the gaps
$(b_{1},a_{2}),...,(b_{q-1},a_{q}),(b_{q},a_{1})$ of the support
of the measure $\nu $, the last gap goes through the infinity
\cite{Ap:84}. For
another form of $\mathcal{D}_{\sigma },\mathcal{G}_{\sigma },\mathcal{R}%
_{\sigma }$, and $\mathcal{S}_{\sigma }$ see \cite{So-Yu:97,Pe-Yu:03}.

The case $q=1$ in (\ref{sigs}) of polynomials orthogonal on a single
interval dates back to S. Bernstein, Szeg\"{o} and Akhiezer \cite{Sz:75}.

The components of the vector $\alpha =\{\alpha _{l}\}_{l=1}^{q-1}$ are
rationally independent generically in $\sigma $, thus the sequences $\{%
\mathcal{D}_{\sigma }(\lambda ,n\alpha )\}_{n\in \mathbb{Z}}$, and$\;\{%
\mathcal{G}_{\sigma }(\lambda ,n\alpha )\}_{n\in \mathbb{Z}}$ for any fixed $%
\lambda $ and the sequences $\;\{\mathcal{R}_{\sigma }(n\alpha
)\}_{n\in \mathbb{Z}}$, and$\;\{\mathcal{S}_{\sigma }(n\alpha
)\}_{n\in \mathbb{Z}}$ are quasiperiodic in $n$ (see
\cite{Ap:84,Du-Co:90,Le:87,So-Yu:97}). As an early precursor of this
fact we mention a result by Akhiezer \cite{Ak:33}, according to
which if $\sigma $ consists of two intervals, then a certain
characteristic of corresponding extremal polynomials of degree $n$
can be expressed via the Jacobi elliptic functions as $n\rightarrow
\infty $. As a result the characteristic does not converge as
$n\rightarrow \infty $ but has a set of limit points that fill a
specific interval generically in the intervals lengths in
(\ref{sigs}).

\subsection{Orthogonal polynomials with respect to varying weights}

Let $V:\mathbb{R}\rightarrow \mathbb{R}_{+}$ be real analytic and such that%
\begin{equation}
\lim_{|\lambda |\rightarrow \infty }\left. V(\lambda )\right/ \log (\lambda
^{2}+1)=\infty .  \label{Vlog}
\end{equation}%
Consider orthonormal polynomials (\ref{wn}) -- (\ref{psi}). To
describe their asymptotics we introduce the functional (cf
(\ref{Es})):

\begin{equation}
\mathcal{E}_{V}[m]=-\int \int \log |\lambda -\mu |m(d\lambda )m(d\mu
)+\int V(\lambda )m(d\lambda ),  \label{EV}
\end{equation}%
defined on the set $\mathcal{M}_1(\mathbb{R})$ of non-negative
unit measures on $\mathbb{R}$.

The functional (\ref{EV}) possesses a unique minimizer $N$%
\begin{equation}\label{mEN}
\min_{m\in \mathcal{M}_{1}(\mathbb{R})}\mathcal{E}_{V}[m]=\mathcal{E}%
_{V}[N ].
\end{equation}%
The variational problem, defined by (\ref{EV}), goes back to Gauss
and is called the minimum energy problem in the external field $V$
(see recent book \cite{Sa-To:97} for a rather complete account of
results and references concerning the problem). The unit measure $N$
minimizing (\ref{EV}) is called the equilibrium measure in the
external field $V$ because of its evident electrostatic
interpretation as the equilibrium distribution of linear charges on
the ideal conductor occupying the axis $\mathbb{R}$ and confined by
the external electric field of potential $V$. We stress that the
corresponding variational problem determines both the (compact)
support $\sigma $ of the measure and the form
of the measure.

The problem is equivalent to the relations \cite{Sa-To:97}%
\begin{equation}
\Phi (\lambda )=-l_{V},\;\lambda \in \sigma ,  \label{EL1}
\end{equation}%
\begin{equation}
\Phi (\lambda )\geq -l_{V},\;\lambda \in \mathbb{R\setminus }\sigma ,
\label{EL2}
\end{equation}%
where%
\begin{equation}
\Phi (\lambda )=V(\lambda )-2\int_{\sigma }\log |\lambda -\mu |N(d\mu ).
\label{Phi}
\end{equation}%
This should be compared with the variational problem (\ref{Es}) of
potential theory, where the external field is absent but the support
$\sigma $ is given. This problem can be viewed as a particular case
of (\ref{EV}), corresponding to a sequence of potentials approaching
$V=\chi _{\sigma }^{-1}-1$, where $\chi _{\sigma }$ is the indicator of $%
\sigma $.

The minimum energy problem in the external field  arises in
various domains of analysis and its applications \cite%
{De-Co:98,Sa-To:97,To:94}. We will use here a link with Random
Matrix Theory. It was argued by Wigner in the 50th (see
\cite{Me:92} for references and discussions), and is shown in
\cite{BPS:95,Jo:98} that the measure $\overline{N}_{n}$ of
(\ref{Nbar}) converges weakly as $n\rightarrow \infty $ to the
unique minimizer $N$ in (\ref{mEN}). Moreover, the random measure
(\ref{NCM}) converges weakly to $N$ with probability 1 as
$n\rightarrow \infty $.

Assume that $V$ is such that the support of $N$ is a union of $q$ disjoint
intervals as in (\ref{sigs}). Introduce the non-increasing function (cf (\ref%
{nu}))
\begin{equation}
N(\lambda )=N((\lambda ,\infty )),  \label{IDS}
\end{equation}%
and the $(q-1)$ - dimensional vector (cf (\ref{alpha}))%
\begin{equation}
\beta =\{\beta _{l}\}_{l=1}^{q-1},\ \beta _{l}=N(a_{l+1}).  \label{beta}
\end{equation}%
With this notation the asymptotics for orthogonal polynomials with varying
weight found in \cite{De-Co:99}, Theorem 1.1, can be written as follows.
There exist continuous functions $\mathcal{D}_{V}:\sigma \times
T^{q-1}\rightarrow \mathbb{R}_{+}$, and $\mathcal{G}_{V}:\sigma \times
T^{q-1}\rightarrow \mathbb{R}$, and $0<\tau \leq 1$ such that if $\lambda $
belongs to the interior of the support $\sigma $ (\ref{sigs}) of $N$, we
have (cf (\ref{psis})):
\begin{eqnarray}\label{psiV}
\psi _{n}^{(n)}(\lambda )&=&(2\mathcal{D}_{V}(\lambda ,n\beta ))^{1/2}
\\ &\times& \cos %
\Big(\pi nN(\lambda )+\mathcal{G}_{V}(\lambda ,n\beta
)\Big)+O(n^{-\tau }),\ n\rightarrow \infty , \nonumber
\end{eqnarray}%
where $\psi _{l}^{(n)}(\lambda )$ is defined in (\ref{psi}), and $n\beta
=(n\beta _{1},...,n\beta _{q-1})$. If $\lambda $ belongs to the exterior of $%
\sigma $, then $\psi _{n}^{(n)}$ decays exponentially in $n$ as $%
n\rightarrow \infty $.

Similar asymptotic formulas are valid for coefficients of the Jacobi matrix $%
J^{(n)}$ of (\ref{Jn}). Namely, according to \cite{De-Co:99}, formula
(1.64), there exist continuous functions $\mathcal{R}_{V}:\mathbb{T}%
^{q-1}\rightarrow \mathbb{R}_{+}$ and $\mathcal{S}_{V}:\mathbb{T}%
^{q-1}\rightarrow \mathbb{R}$ such that we have (cf (\ref{rsRSs}))
\begin{equation}
r_{n}^{(n)}=\mathcal{R}_{V}(n\beta )+O(n^{-\tau }),\;s_{n}^{(n)}=\mathcal{S}%
_{V}(n\beta )+O(n^{-\tau }),\;n\rightarrow \infty .  \label{rsRSV}
\end{equation}%
It will be argued below that the functions $\mathcal{D}_{V},\mathcal{G}_{V},%
\mathcal{R}_{V}$, and $\mathcal{S}_{V}$ differ from the functions $\mathcal{D%
}_{\sigma },\mathcal{G}_{\sigma },\mathcal{R}_{\sigma }$ and $\mathcal{S}%
_{\sigma }$ of formulas (\ref{dgan}) and (\ref{rsRSs}) of the previous
subsection only by a shift in the argument. Hence the main difference in
asymptotic formulas of the previous and this subsection is that in the
former the "rotation number" and the frequencies are determined by the
measure $\nu $ (see (\ref{nu}) -- (\ref{alpha})), minimizing the functional (%
\ref{Es}), while in the latter these quantities (see (\ref{IDS}) -- (\ref%
{beta})) are determined by the measure $N$, minimizing the functional (\ref%
{EV}).

\section{Quasiperiodic Jacobi matrices}

\subsection{Ordinary orthogonal polynomials}

Consider orthonormal polynomials with respect to the weight, whose support
is a union of $q$ disjoint intervals (\ref{sigs}). Denote $J_{\sigma }$ the
semi-infinite Jacobi matrix, associated with the polynomials and let $%
\{r_{l},s_{l}\}_{l\geq 0}$ be the non-zero coefficients of
$J_{\sigma }$ (see (\ref{ur}) -- (\ref{Jn}), in which the
super-index $(n)$ is omitted). Introduce the double-infinite
Jacobi matrix $J_{\sigma ,n}$, setting
\begin{equation}
r_{k,n}=\left\{
\begin{array}{cc}
r_{k+n}, & k\geq -n, \\
0, & k<-n,%
\end{array}%
\right. \qquad s_{l,n}=\left\{
\begin{array}{cc}
s_{k+n}, & k\geq -n, \\
0, & k<-n.%
\end{array}%
\right.  \label{rsnn}
\end{equation}%
We will denote by the same symbol $J_{\sigma ,n}$ the selfadjoint operator
in $l^{2}(\mathbb{Z})$, defined by the matrix.

Now the asymptotics (\ref{rsRSs}) allow us to define a family of "limiting"
double-infinite Jacobi matrices and corresponding selfadjoint operators in $%
l^{2}(\mathbb{Z})$. Assume for the sake of definiteness that the components
of the $(q-1)$-dimensional vector $\alpha $ of (\ref{alpha}) are rationally
independent. Then for any $x=(x_{1},...,x_{q-1})\in \mathbb{T}^{q-1}$ there
exists a subsequence $\{n_{i}(x)\}_{i\geq 1}$, such that
\begin{equation}
\lim_{i\rightarrow \infty }\{n_{i}(x)\alpha
_{l}\}=x_{l},\;l=1,...,q-1, \label{alim}
\end{equation}%
where $\{t\}$ denotes the fractional part of $t\in \mathbb{R}$. This and (%
\ref{rsRSs}) imply that for any $k\in \mathbb{Z}$ we have
\begin{equation}
\lim_{i\rightarrow \infty
}r_{n_{i}(x)+k}=\mathcal{R}_{\sigma}(k\alpha
+x),\;\lim_{i\rightarrow \infty
}s_{n_{i}(x)+k}=\mathcal{S}_{\sigma}(k\alpha +x). \label{rsls}
\end{equation}%
In other words the sequence $\{J_{\sigma ,n_{i}(x)}\}_{i\geq 1}$
of selfadjoint operators, defined in $l^{2}(\mathbb{Z})$ by the
double infinite Jacobi matrices with coefficients (\ref{rsnn}),
converges strongly to the operator in $l^{2}(\mathbb{Z})$, defined
by the double - infinite Jacobi matrix $J_{\sigma }(x) $ with
coefficients
\begin{equation}
\mathcal{R}_{\sigma }(k\alpha +x),\quad \mathcal{S}_{\sigma
}(k\alpha +x),\;k\in \mathbb{Z.}  \label{Js}
\end{equation}%
The matrices $J_{\sigma }(x) $ arise in spectral theory and
integrable systems \cite{Du-Co:90,Le:87,Te:99} and is known there
as finite band Jacobi matrices.

Write the three-term recursion relation for $J_{\sigma ,n}$:%
\[
r_{n+k}\psi _{n+k+1}+s_{n+k}\psi _{n+k}+r_{n+k-1}\psi
_{n+k-1}=\lambda \psi _{n+k},\;k\geq -n,\;\lambda \in \sigma .
\]%
Setting here $n=n_{i}(x)$, using asymptotics (\ref{psis}), and
taking into account that in the obtained asymptotic equality the
coefficients in front of "fast oscillating" expressions $ \cos
(\pi n_{j}(x)\nu (\lambda ))$ and $\sin (\pi n_{j}(x)\nu (\lambda
))$ should be both zero at the limit $i\rightarrow \infty $, we
find that for any $\lambda $, belonging to the interior
of $\sigma $, the sequences%
\begin{equation}
\{(\mathcal{D}_{\sigma }(\lambda ,k\alpha +x))^{1/2}\cos (\pi \nu
(\lambda )k+\mathcal{G}_{\sigma}(k\alpha +x))\}_{k\in \mathbb{Z}},
\label{pc}
\end{equation}%
and%
\begin{equation}
\{(\mathcal{D}_{\sigma }(\lambda ,k\alpha +x))^{1/2}\sin (\pi \nu
(\lambda )k+\mathcal{G}_{\sigma }(k\alpha +x))\}_{k\in \mathbb{Z}}
\label{ps}
\end{equation}%
satisfy the limiting three term recurrence relations, defined by
the coefficients (\ref{Js}). In other words, sequences (\ref{pc})
and (\ref{ps}) are generalized eigenfunctions of $J_{\sigma }(x)$
for every $\lambda$, belonging to the interior of $\sigma $.

Note now that by general principles \cite{Ak:62}
the resolution of identity $\mathcal{E}%
_{J_{\sigma }}$ of the initial Jacobi matrix $J_{\sigma }$ is
\begin{equation}
(\mathcal{E}_{J_{\sigma }}(d\lambda ))_{jk}=\psi _{j}(\lambda )\psi
_{k}(\lambda )d\lambda ,\;j,k\geq 0,  \label{EJs}
\end{equation}%
in particular
\begin{equation}\label{Ede}
\int_{\sigma}(\mathcal{E}_{J_{\sigma }}(d\lambda
))_{jk}=\delta_{jk}.
\end{equation}

Hence, the resolution of identity $\mathcal{E}_{J_{\sigma ,n}}$ of $%
J_{\sigma ,n}$, defined by (\ref{rsnn}), is%
\[
(\mathcal{E}_{J_{\sigma ,n}}(d\lambda ))_{jk}=\left\{
\begin{array}{cc}
\psi _{n+j}(\lambda )\psi _{n+k}(\lambda )d\lambda , & j,k\geq -n \\
0, & \mathrm{otherwise}.%
\end{array}%
\right.
\]%
This and asymptotics (\ref{psis}) yield for the weak
limit of the above
projection-valued measure, the resolution of identity $\mathcal{E}%
_{J_{\sigma }(x)}$ of $J_{\sigma }(x)$:%
\begin{eqnarray}
&&(\mathcal{E}_{J_{\sigma }(x)}(d\lambda ))_{jk} =\chi _{\sigma }(\lambda )(%
\mathcal{D}_{\sigma }(\lambda ,j\alpha +x)\mathcal{D}_{\sigma
}(\lambda ,k\alpha +x))^{1/2}  \label{EJsx} \\ &&\hspace{1cm}
\times \cos \Big( \pi \nu (\lambda )(j-k)+\mathcal{G}_{\sigma }(j\alpha +x)-%
\mathcal{G}_{\sigma }(k\alpha +x)\Big) d\lambda ,\;j,k\in
\mathbb{Z}, \nonumber
\end{eqnarray}%
where $\chi _{\sigma }$ is the indicator of $\sigma $. Denoting $\varphi
_{j}(x)=\pi \nu (\lambda )j+\mathcal{G}(j\alpha +x),\;j\in \mathbb{Z}$, we
can write the cosine above as $\cos \varphi _{j}(x)\cos \varphi _{k}(x)+\sin
\varphi _{j}(x)\sin \varphi _{k}(x)$. This shows that the r.h.s. of (\ref%
{EJsx}) is the linear combination of (\ref{pc}) -- (\ref{ps}). Besides, the
equality%
\[
\int_{\sigma }(\mathcal{E}_{J_{\sigma }(x)}(d\lambda
))_{jk}=\delta _{jk},\;j,k\in \mathbb{Z}
\]%
that can also be proved by the limiting transition
$n_{i}(x)\rightarrow \infty $ in (\ref{Ede}), implies that the union
of the sequences (\ref{pc}) -- (\ref{ps}) for
all $\lambda $ of the interior of $\sigma $ forms a complete system in $%
l^{2}(\mathbb{Z})$.

Introducing
\begin{equation}
\Psi _{j}(\lambda ,x)=e^{i\pi \nu (\lambda )j}u_{j}(\lambda ,x),  \label{qB1}
\end{equation}%
where
\begin{equation}
u_{j}(\lambda ,x)=\mathcal{U}(\lambda ,j\alpha +x),\;\mathcal{U}(\lambda ,x)=%
\mathcal{D}_{\sigma }^{1/2}(\lambda ,x)e^{i\mathcal{G}_{\sigma
}(\lambda ,x)}, \label{qB2}
\end{equation}%
we conclude from the above that the union of sequences
\begin{equation}
\{\Psi _{j}(\lambda ,x)\}_{j\in \mathbb{Z}},\;\{\overline{\Psi _{j}(\lambda
,x)}\}_{j\in \mathbb{Z}}  \label{qB}
\end{equation}%
for all $\lambda $ of the interior of $\sigma $ also forms a complete system
of generalized eigenfunctions of the "limiting" selfadjoint operator $%
J_{\sigma }(x)$, acting in $l^{2}(\mathbb{Z})$. The system is known in
spectral theory as the quasi-Bloch generalized eigenfunctions, because in
the case of periodic coefficients (see e.g. (\ref{abper})) they are well
known Floquet-Bloch solutions of corresponding finite-difference equation.
In this context $\nu (\lambda )$ is called the quasi-momentum as the
function of spectral parameter.

Recall now that if $A=\{A(x)\}_{x\in \mathbb{T}^{q-1}}$ is a selfadjoint
quasiperiodic operator in $l^{2}(\mathbb{Z})$, then its Integrated Density
of States can be defined as
\begin{equation}
k_{A}(\lambda
)=\int_{\mathbb{T}^{q-1}}(\mathcal{E}_{A(x)}((\lambda ,\infty
))_{00}\: dx,  \label{IDSA}
\end{equation}%
where $\mathcal{E}_{A(x)}$ is the resolution of identity of $A(x)$ (see \cite%
{Cy-Co:87,Pa-Fi:92} for this and a more general case of ergodic operators).

By using the above definition and (\ref{EJs}), we find%
\begin{equation}
k_{J_{\sigma }(\cdot )}(d\lambda )=\left( \int_{\mathbb{T}^{q-1}}\mathcal{D}%
_{\sigma }(\lambda ,x)dx\right) d\lambda,  \label{knus}
\end{equation}%
where $k_{J_{\sigma }(\cdot )}(d\lambda )$ is the measure,
corresponding to the non-increasing function $k_{J_{\sigma }(\cdot
)}(\lambda )$ in (\ref{IDSA}).

Another definition of the Integrated Density of States is as
follows. Consider the restriction $A_{n}(x)$ of $A(x)$ to a finite
interval $[1,n]$, imposing certain selfadjoint boundary conditions
at the endpoints of the interval. The spectrum of $A_{n}(x)$ is a
finite set and we can introduce its Normalizes Counting Measure of
eigenvalues $k_{n}$ as the divided by $n$ number of
eigenvalues of $A_{n}(x)$ in the interval $(\lambda ,\infty )$ (cf (\ref{NCM}%
)). It is known (see e.g. \cite{Cy-Co:87,Pa-Fi:92}) that $k_{n}$ converges
weakly to (\ref{IDSA}) for any $x\in \mathbb{T}^{q-1}$.

In the case of operators $J_{\sigma }(x)$, possessing the complete family of
quasi Bloch generalized eigenfunctions (\ref{pc}) -- (\ref{ps}), it can be
shown that $k_{J_{\sigma }(\cdot )}=\nu $, i.e. that the Integrated Density
of States of $J_{\sigma }(\cdot )$ coincides with its quasi-momentum as the
function of the spectral parameter, and we have from (\ref{knus})%
\begin{eqnarray}\label{knu}
k_{J_{\sigma }(\cdot )}(d\lambda )&=&\nu (d\lambda )\\
&=&\left( \int_{\mathbb{T}%
^{q-1}}\mathcal{D}_{\sigma }(\lambda ,x)dx\right) d\lambda .
\nonumber
\end{eqnarray}%
We found, in particular, a relation between two quantities of
asymptotics (\ref{psis}).

Another important characteristics of quasi periodic (more generally,
ergodic) operators is the Lyapunov exponent $\gamma_{A} (\lambda )$,
defined as the rate of exponential growth of the Cauchy solutions of
the corresponding finite difference equation of second order. The
Lyapunov exponent and the Integrated Density of States are related
by the Thouless formula (see e.g. \cite{Pa-Fi:92}, formula (11.82)).
In the case of the quasiperiodic Jacobi matrix $J_{\sigma}(x)$ the
formula is
\begin{equation}
\gamma_{J_{\sigma}(\cdot)} (\lambda
)=-\int_{\mathbb{T}^{q-1}}\mathcal{R}_{\sigma }(x)dx+\int_{\sigma
}\log |\lambda -\mu |k_{J_{\sigma}(\cdot)}(d\lambda ).  \label{Tho}
\end{equation}%
Since the generalized functions of $J_{\sigma }(x)$ are bounded and
do not
decay at infinity (see (\ref{pc}) -- (\ref{ps}) or (\ref{qB1}) -- (\ref{qB2}%
)), we have%
\[
\gamma_{J_{\sigma}(\cdot)}(\lambda )=0,\;\lambda \in \sigma .
\]%
Hence, the l.h.s. of
(\ref{Tho}) is zero if $\lambda \in \sigma $. In view
of (\ref{knus}) the obtained relation is just the Euler-Lagrange equation (%
\ref{Es1}) for the functional (\ref{Es}).


\subsection{Orthogonal polynomials with respect to varying weights}

We will present here constructions, similar to those of the previous
subsection but for orthogonal polynomials with respect to varying weights.
To this end it is useful to make explicit the amplitude of the potential $V$
in (\ref{wn}), i.e. to replace $V$ by $V/g,\;g>0$. In what follows we will
keep $V$ fixed and vary $g$. Thus orthonormal polynomials (\ref{ortho}) and
related quantities will depend on $g$. To control this dependence we will
use results of papers \cite{Bu-Ra:99,Ku-Mc:00}.

Note first that if the potential is real analytic, then the minimizer $N$ of
(\ref{EV}) possesses a density $\rho $ supported on a finite union of finite
intervals $\sigma $ \cite{BPS:95,De-Co:98}.

According to \cite{De-Co:99} asymptotics (\ref{psiV}) --
(\ref{rsRSV}) are most precise and well behaving if a real analytic
potential, satisfying  (\ref{Vlog}), is regular (see
\cite{De-Co:99}, formulas (1.12) and (1.13)).  This condition
implies, in particular, that the density of the measure $N$ in
(\ref{mEN}) is strictly positive on the interior of its support
$\sigma $, and vanishes as a square root at each edge of $%
\sigma $. Furthermore, following \cite{Ku-Mc:00}, we say that $g$ is
regular for $V$ if $V/g$ is a regular potential. If $g_{0}$ is
regular for $V$, and
\[
\sigma
_{g_{0}}=\bigcup\limits_{l=1}^{q}[a_{l}(g_{0}),b_{l}(g_{0})]
\]%
is the support of the equilibrium measure $N_{g_{0}}$corresponding to $%
V/g_{0}$, then there exists an open neighborhood of $g_{0}$, consisting of
regular values $g$ for $V$ and
\begin{equation}
\sigma _{g}=\bigcup\limits_{l=1}^{q}[a_{l}(g),b_{l}(g)]  \label{sg}
\end{equation}%
with the same number $q$ of intervals. Besides, $a_{l}$ and $b_{l}$
are real analytic, $a_{l}$ is strictly decreasing and $b_{l}$ is
strictly increasing in $g$.

We will also need the following formula, relating $N_{g}$ and $\nu
_{g}$, minimizing correspondingly (\ref{EV}) with $V/g$ instead
$V$ and (\ref{Es})
with $\sigma _{g}$ instead of $\sigma $ \cite{Bu-Ra:99}:%
\begin{equation}
N_{g}=g^{-1}\int_{0}^{g}\nu _{g^{\prime }}dg^{\prime }.  \label{Nnu}
\end{equation}%
The formula was proved in \cite{Bu-Ra:99} in a fairly general
setting, including piece-wise continuous $V$'s. Its particular cases
are given in \cite{Pa:96,Bu-Pa:02}, where its spectral and
asymptotic meaning made explicit, related to a kind of "adiabatic"
regime in $g$ for corresponding Jacobi matrix (\ref{Jn}) (see also
the derivation of formula (\ref{rD0}) below).

Now we can give an analog of constructions of previous subsection, i.e. the
"limiting" Jacobi matrix with quasiperiodic coefficients. We confine
ourselves again to the case of rationally independent components of vector $%
\beta $ of (\ref{beta}), a generic case in $g$.

Consider the coefficients $r_{l}^{(n)}$ of the Jacobi matrix
(\ref{Jn}), associated with orthonormal polynomials
$\{p_{l}^{(n)}\}_{l\geq 0}$ with varying weight. Introducing
explicitly the dependence of coefficients on $g$
and writing in view of (\ref{wn}) with $%
V/g$ instead of $V$
\begin{equation}\label{Vgln}
n \frac{V}{g} = l \frac{V}{gl/n},
\end{equation}
we get
\begin{equation}
r_{l}^{(n)}(g)=r_{l}^{(l)}(gl/n).  \label{rlng}
\end{equation}%
Setting here $l=n_{i}(x)+k$, where now (cf (\ref{alim}))
\begin{equation}
\lim_{i\rightarrow \infty }\{n_{i}(x)\beta
_{l}\}=x_{l},\;l=1,..,q-1, \label{blim}
\end{equation}%
$x=\{x_{l}\}_{l=1}^{q-1}$ is a point of $\mathbb{T}^{q-1}$, and
$k$ is an arbitrary fixed integer, we obtain in view of
(\ref{rsRSV}) and the continuity of $\mathcal{R}_V$ in $g$ and
$x$,
and of $\beta$ in $g$:%
\begin{eqnarray}
\lim_{i\rightarrow \infty }r_{n_{i}(x)+k}^{(n_{i}(x))}(g)
&=&\lim_{i \to \infty} \mathcal{R}%
_{V}\left( \frac{n_{i}(x)+k}{n_{i}(x)}g,(n_{i}(x)+k)\beta
\left( \frac{n_{i}(x)+k}{n_{i}(x)}%
g\right) \right)   \label{rlim} \\
&=&\mathcal{R}_{V}\left( g,k\widetilde{\alpha }(g)+x\right) .
\nonumber
\end{eqnarray}%
where%
\begin{equation}
\widetilde{\alpha }(g)=(g\beta (g))^{\prime }.  \label{atbd}
\end{equation}%
Analogous relation is valid for the diagonal entries of $J^{(n)}$:%
\[
\lim_{i\rightarrow \infty }s_{n_{i}(x)+k}^{(n_{i}(x))}(g)=\mathcal{S}_{V}\left( g,k%
\widetilde{\alpha }(g)+x\right) .
\]%
Now, by using formula (\ref{Nnu}), we find an important relation%
\begin{equation}
\widetilde{\alpha }(g)=\alpha (g),  \label{ata}
\end{equation}%
where $\alpha (g)$ is defined by (\ref{alpha}) with $\nu _{g}$ instead of $%
\nu $. We conclude from the above that limiting coefficients are (cf
(\ref{rsls}))
\begin{eqnarray}\label{rslim}
&&\lim_{i\rightarrow
\infty}r_{n_{i}(x)+k}^{(n_{i}(x))}(g)=\mathcal{R}_{V}\left(
g,k\alpha (g)+x\right), \;k\in \mathbb{Z}, \\&& \lim_{i\rightarrow
\infty }s_{n_{i}(x)+k}^{(n_{i}(x))}(g)=\mathcal{S}_{V}\left(
g,k\alpha (g)+x\right), \;k\in \mathbb{Z}.  \nonumber
\end{eqnarray}%
As a result we obtain a quasiperiodic Jacobi matrix $J_{V/g}(x)$, defined by
the coefficients (cf (\ref{Js}))%
\begin{equation}
\mathcal{R}_{V}\left( g,k\alpha (g)+x\right), \;
\mathcal{S}_{V}\left( g,k\alpha (g)+x\right), \; k\in \mathbb{Z},
\label{JV}
\end{equation}%
and having the frequencies $(\alpha _{1}(g),...,\alpha _{q-1}(g))$, obtained
from (\ref{nu}) -- (\ref{alpha}) in which $\nu $ is $\nu _{g}$, hence $%
\sigma $ is the support $\sigma _{g}$ of $N_{g}$. Note that $J_{V/g}(x)$ is
the limit in the sense (\ref{blim}) in the strong operator topology of $%
l^{2}(\mathbb{Z})$ of matrices $J_{V/g,n}^{(n)}$, whose coefficients are
defined by formulas (\ref{rsnn}) with $r_{n+k}^{(n)}$ and $s_{n+k}^{(n)}$
instead of $r_{n+k}$ and $s_{n+k}$.

Applying the same limiting argument to asymptotic formula (\ref{psiV}), and
by using (\ref{blim}) and (\ref{Nnu}), we obtain for any fixed $k\in \mathbb{%
Z}$%
\begin{eqnarray}\label{psiVg}
\psi _{n_{i}(x)+k}^{(n)}(\lambda ) &=&(2\mathcal{D}_{V}(\lambda
,g,k\alpha(g)+x))^{1/2} \cos \Big(\pi n_{i}(x)N_{g}(\lambda )\\
&+&\pi k\nu _{g}(\lambda )+\mathcal{%
G}_{V}(\lambda ,g,k\alpha (g)+x)\Big)+o(1),\ n_{i}(x)\rightarrow
\infty. \nonumber
\end{eqnarray}%
By using these formulas, the exponential decay of $\psi _{n+k}^{(n)}(\lambda
)$ outside $\sigma _{g},$ and the limit (\ref{blim}), we obtain the complete
system of generalized eigenfunctions and the resolution of identity $%
\mathcal{E}_{J_{V/g}(x)}$ of $J_{V/g}(x)$, given by (\ref{pc}) -- (\ref{ps})
(or (\ref{qB})) and (\ref{EJs}), in which subindex $\sigma $ is replaced by
the subindex $V/g$. In particular, we have for the diagonal entries of $%
\mathcal{E}_{J_{V/g}(x)}$:
\begin{equation}
\left( \mathcal{E}_{J_{V/g}(x)}\right) _{kk}(d\lambda )=\chi _{\sigma
_{g}}(\lambda )\mathcal{D}_{V}(\lambda ,g,k\alpha (g)+x),\;k\in \mathbb{Z},
\label{EVjk}
\end{equation}%
where $\chi _{\sigma }$ is the indicator of $\sigma $ (cf
(\ref{EJsx}) with $j=k$). The limit here is the weak limit of
measures. The support in $\lambda $ of the r.h.s. of this formula is
the support $\sigma _{g}$ of the equilibrium measure $N_{g}$.
This implies that the spectrum of the quasiperiodic matrix $J_{V/g}(x)$ is $%
\sigma _{g}$. Note that the spectrum of the "initial" double
infinite matrix $J_{n}^{(n)}$, defined analogously (\ref{rsnn}) but
via $r^{(n)}_{n+k}$ and $s^{(n)}_{n+k}$, is $\mathbb{R}$ for all $n
< \infty$.

Besides, arguing as in obtaining (\ref{knus}), we find for the
Integrated
Density of States measure $k_{J_{V/g}(\cdot)}$ of $J_{V/g}(x)$:%
\begin{eqnarray}\label{knuV}
k_{J_{V/g}(\cdot )}(d\lambda )&=&\nu _{g}(d\lambda )\\&=&\left( \int_{\mathbb{T}%
^{q-1}}\mathcal{D}_{V}(\lambda ,g,x)dx\right) d\lambda . \nonumber
\end{eqnarray}%
Comparing the first equality of this formula with the first equality
of (\ref{knu}) in which $\sigma$ is replaced by $\sigma_g$, we
conclude that $J_{\sigma _{g}}(x)$ and $J_{V/g}(x)$ have the same
spectrum and the same Integrated Density of States.

The coincidence of spectra of $J_{\sigma _{g}}(x)$ and
$J_{V/g}(x)$ implies (see \cite{Du-Co:90,Le:87,Te:99}) that each
of them is an isospectral deformation of another, i.e. that the
coefficients (\ref{Js}) of $J_{\sigma _{g}}(x)$ differ from the
coefficients (\ref{JV}) of $J_{V/g}(x)$ just by a shift of their
argument. This fact can also be checked directly, by
comparing explicit formulas for both sets of coefficients, given \cite%
{Ap:84,Du-Co:90,Te:99} and in \cite{De-Co:99} correspondingly, and
by using again the trick with infinitesimal variation of the
amplitude of potential (see Appendix).

Here is one more link between two classes of polynomials and
spectral theory. It concerns the Lyapunov exponents of $J_{\sigma
_{g}}(x)$ and $J_{V/g}(x)$ and the potential. It can be shown
\cite{Bu-Ra:99} that the Lyapunov exponents of the both matrices
coincide and if $\gamma _{g}(\lambda )$ is their common value, then
\begin{equation}
V(\lambda )=2\int_{0}^{g}\gamma _{g^{\prime }}(\lambda )dg^{\prime
},\;\lambda \in \sigma _{g}.  \label{Vga}
\end{equation}


\subsection{Periodic Jacobi matrices}

Here we consider a class of polynomial potentials $V$ in (\ref{wn}) for
which corresponding Jacobi matrices $J_{\sigma _{g}}(x)$ and $J_{V/g}(x)$
have periodic coefficients. Besides, several quantities related to the
matrices and orthogonal polynomials can be found explicitly. We follow \cite%
{Bu-Pa:02}.

Let $v$ be a polynomial of degree $q$ with real coefficients and with the
leading term $z^{q}$. Assume that there exists $g>0$ such that all zeros of
the polynomial $v^{2}-4g$ are real and simple and set
\begin{equation}
V(\lambda )=\frac{v^{2}(\lambda )}{2q}.  \label{Vv}
\end{equation}%
We will show that in this case coefficients of $J_{\sigma _{g}}(x)$ and $%
J_{V/g}(x)$ are $q$-periodic and their spectrum is
\begin{equation}
\sigma _{g}=\{\lambda :v^{2}(\lambda )-4g\leq 0\}.  \label{ssg}
\end{equation}%
We show first that the equilibrium measures $N_{g}$ and $\nu _{g}$ for (\ref%
{EV}) and (\ref{Es}) are:%
\begin{equation}
N_{g}(d\lambda )=\rho _{g}(\lambda )d\lambda ,\;\rho _{g}(\lambda )=\frac{%
|v^{\prime }(\lambda )|}{2\pi gq}|v^{2}(\lambda )-4g|^{1/2}\chi _{\sigma
_{g}}(\lambda ),  \label{rot}
\end{equation}%
and%
\begin{equation}
\nu _{g}(d\lambda )=d_{g}(\lambda )d\lambda ,\;d_{g}(\lambda )=\frac{%
|v^{\prime }(\lambda )|}{\pi q}|v^{2}(\lambda )-4g|^{-1/2}\chi _{\sigma
_{g}}(\lambda ).  \label{dot}
\end{equation}%
Indeed, it is a matter of direct calculations to find that if
$\rho_g$ is given by (\ref{rot}), then
\begin{eqnarray}
w_{V/g}(z)  :&=&-\int_{\sigma _{g}}\log (z-\mu )\rho _{g}(\mu )d\mu
\label{wg}=\frac{1}{q}\Big[ u(z)\sqrt{u^{2}(z)-1}\\&-&
\log \left( u(z)+\sqrt{u^{2}(z)-1}%
\right) -u^{2}(z)\Big] -(2q)^{-1}\log g/e.  \nonumber
\end{eqnarray}%
where $u=v/2\sqrt{g}$, and we use the branch of logarithm with the cut $%
(-\infty ,0)$ and the argument $\pi $ on the upper edge of the cut and the
branch of $\sqrt{u^{2}-1}$, such that $\sqrt{u^{2}-1}=u+o(1),\;u\rightarrow
\infty $.

On the other hand%
\[
\Re w_{V/g}(\lambda +i0)=-\int_{\sigma _{g}}\log |\lambda -\mu |\rho
_{g}(\mu )d\mu
\]%
is the logarithmic potential of $\,N_{g}(d\lambda )=\rho _{g}(\lambda
)d\lambda $. Now, analyzing the values of $\Phi $ of (\ref{Phi}) in this case%
\begin{equation}
\Phi _{g}(\lambda )=\frac{V(\lambda )}{g}+2\Re w_{V/g}(\lambda +i0)
\label{Phig}
\end{equation}%
with $V$ from (\ref{Vv}), we can check directly the validity of (\ref{EL1})
-- (\ref{EL2}) (with $V/g$ instead of $V$) with the strict inequality in (%
\ref{EL2}) and $l_{V}=(2q)^{-1}\log g/e$. This proves that $N_{g}$ of (\ref%
{rot}) is the minimizer of (\ref{EV}) with $V/g$ instead of $V$.

It can also be proved that $\nu _{g}$ of (\ref{dot}) is the
minimizer of (\ref{Es}) with
$\sigma _{g}$ of (\ref{ssg}) instead of $\sigma $. We can use either (\ref%
{Nnu}) or the above scheme, computing (cf (\ref{wg}))
\begin{eqnarray*}
w_{\sigma _{g}}(z):&=&-\int_{\sigma _{g}}\log (z-\mu )d_{g}(\mu
)d\mu \\&=&-\frac{1}{q}\log \left( u(z)+\sqrt{u^{2}(z)-1}\right)
-\frac{1}{2q}\log g,
\end{eqnarray*}%
and then checking directly (\ref{Es1}) -- (\ref{Es2}).

We will use, however, another argument to prove (\ref{rot}) and
(\ref{dot}). The argument is based on a representation, important
in the inverse problem for periodic operators of second order
\cite{Ma-Os:75,Ma:86}.

Let $u$ be a polynomial of degree $q$ with real coefficients and
such that all zeros of $u^{2}-1$ are real and simple. Then $u$ can
be written in the form
\begin{equation}
u(z)=\cos \theta (z),  \label{uct}
\end{equation}%
in which $\theta (z)$ is the conformal map of the open upper half-plane $%
\mathbb{C}_{+}=\{z\in \mathbf{C}:\mathrm{Im}z>0\}$ onto the domain%
\begin{eqnarray}\label{comb}
\{\theta &:&q_{1}\pi <\Re \theta <q_{2}\pi ,\;\Im \theta
>0\}\\ &&\setminus \bigcup_{q_{1}<l<q_{2},}\{\theta :\Re \theta =l\pi
,\ q_{1}<l<q_{2},\ 0<\Im \theta \leq h_{l}\},  \nonumber
\end{eqnarray}%
Here $q_{1}<q_{2}$ are integers, $q_{2}-q_{1}=q$, $0\leq h_{l}\
<\infty $ and $\theta (\infty )=\infty $. In fact, the r.h.s. of
(\ref{uct}) is a polynomial of degree $q$ if and only if $-\infty
<q_{1}<q_{2}<\infty ,\ q_{2}-q_{1}=q$ \cite{Ma-Os:75}. Function
$\theta (z)$ is analytic in $\mathbb{C}_{+}$ and
continuous in the closed upper half-plane $\overline{\mathbb{C}_{+}}$. When $%
z=\lambda +i0$ varies from $-\infty $ to $\infty $, the limiting value $%
\theta (\lambda +i0)$ runs along the boundary (the "comb") of the domain (%
\ref{comb}), so that either $\Re \theta (\lambda +i0)$ varies from
$(q_{1}+l-1)\pi $ to $(q_{1}+l)\pi $ and $\Im \theta (\lambda
+i0)=0$, if $\lambda $ varies through the $l$th "band" $[a_{l},b_{l}],%
\;l=1,...,q$ of $\sigma _{g}$, or $\Re \theta (\lambda +i0)\equiv 0 \; (\mathrm{%
mod}\pi )$ and $\Im \theta (\lambda +i0)=\kappa ,\;0\leq \kappa \leq h_{l}$,
if $\lambda $ varies through the $l$th "gap" $(b_{l},a_{l+1})$ of $\sigma
_{g}$.

By using the terminology of mathematical physics we can say that $\theta
(z)/\pi q$ is an analytic continuation of the quasimomentum as a function of
energy in the extended band scheme.

We set $u=v/2\sqrt{g}$ in (\ref{uct}). Then the zeros of $u^{2}-1$
are the band edges $-\infty <a_{1}<b_{1}<...<a_{q}<b_{q}, \; \theta
(b_{q})=0, \; q_{1}=-q,\; q_{2}=0$ and $\theta (\lambda +i0)$ varies
from $(-q+l-1)\pi $ to $(-q+l)\pi ,\;l=1,...,q$ when $\lambda $
varies from $a_{l}$ to $b_{l}$ in
the $l$th band. By using (\ref{uct}) and (\ref{comb}) we can rewrite (\ref%
{EL1}) -- (\ref{Phi}), (\ref{Phig}) as
\begin{equation}
\Phi (\lambda )=-\frac{1}{q}\log g/e+\left\{
\begin{array}{rl}
0, & \mbox{if}\ \lambda \in \lbrack a_{l},b_{l}],\;l=1,...,q, \\
g(\kappa _{l}(\lambda )), & \mbox{if}\ \lambda \in \lbrack
b_{l},a_{l+1}],\;l=1,..,q,%
\end{array}%
\right.   \label{2.2}
\end{equation}%
where $a_{l+1}=a_1$,
\[
g(\kappa ):=\frac{2}{q}\left( \frac{\sinh 2\kappa }{2}-\kappa \right) =\frac{%
4}{q}\int_{0}^{\kappa }\sinh ^{2}tdt>0,\ \kappa >0,
\]%
and $\kappa _{l}(\lambda )$ varies from $0$ and $h_{l}>0$, when $\lambda $
varies through the gap $(b_{l},a_{l+1})$. This yields (\ref{EL1}) -- (\ref%
{Phi}), thereby a proof that (\ref{rot}) is the density of the equilibrium
measure $N_{g}$, corresponding to the potential (\ref{Vv}). Moreover, since
the inequality in (\ref{EL2}) is strict in this case, the corresponding
value of $g$ is regular for the potential (\ref{Vv}).

It follows from (\ref{wg}) that
\[
N_{g}(\lambda )=-\pi ^{-1}\Im w_{V/g}(\lambda +i0),
\]%
and then (\ref{uct}) implies that%
\begin{equation}
N_{g}(\lambda )=\frac{1}{\pi q}\left( \theta _{+}(\lambda )-\frac{\sin
2\theta _{+}(\lambda )}{2}\right) ,  \label{Ntet}
\end{equation}%
and similarly%
\begin{equation}
\nu _{g}(\lambda )=\frac{1}{\pi q}\theta _{+}(\lambda ),  \label{nutet}
\end{equation}%
where $\theta _{+}(\lambda )=\Re \theta (\lambda +i0)$, and $\theta
(z)$ is defined in (\ref{uct}). In view of the above properties of
this function, we have
\begin{equation}
N_{g}(a_{l+1})=\nu _{g}(a_{l+1})=\frac{q-l}{q},\quad l=1,...,q-1,
\label{N/q}
\end{equation}%
and then (\ref{alpha}) and (\ref{beta}) imply%
\begin{equation}
\alpha _{l}=\beta _{l}=\frac{q-l}{q},\quad 1=1,...,q-1.  \label{abper}
\end{equation}%
Hence, the coefficients (\ref{Js}) of the matrix $J_{\sigma _{g}}(x)$ and
the coefficients (\ref{JV}) of the matrix $J_{V/g}(x)$ are $q$-periodic in
this case. Moreover, we need not to consider in this case the whole torus $%
\mathbb{T}^{q-1}$ as the set of values of $x$ in (\ref{alim}) and (\ref{blim}%
), but just the set of vertices of the regular $q$-polygon. This
is similar to a standard procedure of the theory of almost
periodic functions, where the corresponding set is the closure of
all limiting points of sequences (\ref{alim}) or (\ref{blim}),
hence depends on arithmetic properties of the frequency vector
$\alpha $ or $\beta$.

Consider simple cases of potentials (\ref{Vv}). The case $q=1$
corresponds
to $v(\lambda )=-\lambda $ and yields%
\begin{eqnarray*}
&&\sigma _{g}=[-2\sqrt{g},2\sqrt{g}],   %
\\
&&\rho _{g}(\lambda )=\frac{1}{2\pi g}(4g-\lambda ^{2})^{1/2}\chi
_{\sigma _{g}}(\lambda ),
\\
&&d_{g}(\lambda )=\frac{1}{\pi }(4g-\lambda ^{2})^{-1/2}\chi
_{\sigma _{g}}(\lambda ).
\end{eqnarray*}%
The first density corresponds to the well known semicircle law by
Wigner for
the Gaussian Unitary Ensemble \cite{Me:92}. The role of polynomials $%
p_{l}^{(n)}$ play $h_{l}(\lambda \sqrt{n/2g})(n/2g)^{1/4}$, where $%
\{h_{l}\}_{l\geq 0}$ are the orthonormal Hermite polynomials. The
second density is the Density of States of the Jacobi matrix with
constant coefficients $r_{l}=\sqrt{g},\;s_{l}=0,\;l\in \mathbb{Z}$.
The matrix  plays here the role of both limiting matrices $J_{\sigma
_{g}}$ and $J_{V/g}$.

The case $q=2$ corresponds to $v(\lambda )=\lambda ^{2}+v_{0},\;v_{0}<-2%
\sqrt{g}$, and yields%
\[
\sigma _{g}=[-b(g),-a(g)]\cup \lbrack a(g),b(g)],
\]
\[a(g)=(|v_{0}|-2\sqrt{g}%
)^{1/2},\;b(g)=(|v_{0}|+2\sqrt{g})^{1/2},
\]%
\[
\rho _{g}(\lambda )=\frac{|\lambda |}{2\pi g}\left( (b^{2}-\lambda
^{2})(\lambda ^{2}-a^{2})\right) ^{1/2}\chi _{\sigma _{g}}(\lambda ),
\]%
\[
d_{g}(\lambda )=\frac{|\lambda |}{\pi }\left( (b^{2}-\lambda ^{2})(\lambda
^{2}-a^{2})\right) ^{-1/2}\chi _{\sigma _{g}}(\lambda ).
\]%
Asymptotics of corresponding orthogonal polynomials were considered in \cite%
{Bl-It:97}. Matrices $J_{\sigma _{g}}$ and $J_{V/g}$ are both of
period 2 and their Density of States is given above.

For a general two interval case, where the corresponding matrices
are quasiperiodic and their coefficients can be expressed via the
Jacobi elliptic functions see \cite{Ak:60,AT:61,Pe:95} (ordinary
polynomials) and \cite {Da-Co:00} (polynomials with varying
weights).

The fact that in the case of ordinary polynomials the limiting finite band
Jacobi matrix is periodic if its spectrum is the inverse image of a
polynomial map (see (\ref{ssg})) is known (see e.g. \cite{Pe:91,Te:99} and
references therein). It is of interest that the same property holds also for
polynomials with varying weights and that the corresponding potential (\ref%
{Vv}) is also polynomial and can be explicitly related to the map.

We mention one more link of asymptotics of orthonormal polynomials
and periodic Jacobi matrices \cite{Bu-Pa:02} that concerns the Hill
discriminant (or the Lyapunov function) of $J_{\sigma _{g}}(x)$ and
$J_{V/g}(x)$ and the polynomial $v$ of (\ref{Vv}). Recall that the
Hill discriminant $\Delta (\lambda )$ is defined as $1/2$ of the
trace of the monodromy (transfer) matrix of corresponding
finite-difference equation of second order with periodic
coefficients and plays an important role in spectral theory (see
e.g. \cite{Ma:86,Si:05}). It can be shown \cite{Bu-Pa:02} that both
matrices have the same Hill discriminant $\Delta _{g}$ and that
\[
\Delta _{g}=v(\lambda )\diagup 2\sqrt{g}.
\]%
We discussed above the case, where the polynomial $v$ in (\ref{Vv})
is such that all zeros of $v^{2}-4g$ are real and simple. Admitting
non-simple (but still real) zeros, we include the case, where two
adjacent bands touch one another or a band is going to appear inside
a gap.

\section{Eigenvalue distribution of random matrices}

Here we discuss briefly certain aspects of eigenvalue distributions of
ensembles (\ref{M}) -- (\ref{dM}), related to the above results, in
particular to the matrix $J_{V(x)}$.

\subsection{Expectation of linear statistics}

According to \cite{BPS:95,Jo:98} we have for any bounded and continuous $\varphi $%
\[
\lim_{n\rightarrow \infty }\mathbf{E}\{N_{n}[\varphi ]\}=\int_{\sigma
}\varphi (\lambda )N(d\lambda ),
\]%
where $N$ is the minimizer of (\ref{EV}). Combining this with (\ref{Nnu}), (%
\ref{EVjk}), and \ref{knuV}), we obtain for the r.h.s. of this formula%
\[
\int_{\sigma }\varphi (\lambda )d\lambda \int_{0}^{1}\nu
_{g}(d\lambda )dg=\int_{0}^{1}dg\int_{\mathbb{T}^{q-1}}(\varphi
(J_{V/g}(x)))_{00} \; dx.
\]%
On the other hand we can always write (\ref{stl}) as%
\[
N_{n}[\varphi ]=n^{-1}\mathrm{Tr}\varphi (M_n),
\]%
and we obtain a kind of "functional correspondence"%
\begin{equation}
\lim_{n\rightarrow \infty }\mathbf{E}\{n^{-1}\mathrm{Tr}\varphi
(M_n)\}=\int_{0}^{1}dg\int_{\mathbb{T}^{q-1}}(\varphi
(J_{V/g}(x)))_{00} \; dx, \label{fcorr}
\end{equation}%
reminiscent to that for ergodic operators, see \cite{Cy-Co:87}, Theorem 9.6,
and \cite{Pa-Fi:92}, Theorem 4.4.

Here is a heuristic argument, explaining the above formula. According to (%
\ref{NbK}) the l.h.s. of the formula includes the orthonormal functions $%
\psi _{l}^{(n)}$ of (\ref{psi}) for $l=0,...,n-1$. Indicating explicitly the
dependence of these functions on $g$ and using the relation (cf (\ref{rlng}%
)):
\begin{equation}
\psi _{l}^{(n)}(\lambda ,g)=\psi _{l}^{(l)}(\lambda ,gl/n),  \label{pnpl}
\end{equation}%
we obtain from (\ref{psiV}) that the leading contribution to $\rho _{n}$ as $%
n\rightarrow \infty $ is
\[
\frac{1}{n}\sum_{l=0}^{n-1}\mathcal{D}_{V}\left( \lambda ,l/n,l\beta
(l/n)\right) ,
\]%
Assuming that $\mathcal{D}_{V}(\lambda ,g,x)$, and $\beta (g)$ are
continuous in $g$, we can say that the summand in this formula is "slow
varying" in $l/n$ and "fast varying" in $l$. This observation results in the
limiting formula for the density $\rho (\lambda )$ of the measure $N$:
\begin{equation}
\rho (\lambda )=\int_{0}^{1}dg\int_{\mathbb{T}^{q-1}}\mathcal{D}_{V}(\lambda
,g,x)dx.  \label{rD0}
\end{equation}%
Using now (\ref{knuV}) and (\ref{Nnu}), we obtain (\ref{fcorr}).

\subsection{Covariance of linear statistics of eigenvalues}

By using (\ref{CDf}) we write (\ref{CovK}) as
\[
\mathbf{Cov}\{N_{n}[\varphi _{1}],N_{n}[\varphi _{2}]\}=\frac{1}{n^{2}}\int
\int \frac{\Delta \varphi _{1}}{\Delta \lambda }\frac{\Delta \varphi _{2}}{%
\Delta \lambda }\mathbf{C}_{n}(\lambda _{1},\lambda _{2})d\lambda
_{1}d\lambda _{2},
\]%
where $\Delta \varphi /\Delta \lambda $ is defined in (\ref{dede}),%
\[
\mathbf{C}_{n}(\lambda _{1},\lambda _{2})=(r_{n-1}^{(n)})^{2}\left(
e_{n,n}^{(n)}(\lambda _{1})e_{n-1,n-1}^{(n)}(\lambda
_{2})-e_{n,n-1}^{(n)}(\lambda _{1})e_{n,n-1}^{(n)}(\lambda _{2})\right)
\]%
and
\[e_{l,m}^{(n)}(\lambda )=\psi _{l}^{(n)}(\lambda )\psi
_{m}^{(n)}(\lambda )
\]
is the density of the resolution of identity $\{\mathcal{E}%
_{J^{(n)}}(d\lambda )\}_{l,m=0}^\infty$ of matrix (\ref{Jn}) (cf
(\ref{EJs})). Thus, assuming that $\varphi _{1,2}$ are bounded and
of the class $C^{1}$ and passing to a subsequence
$\{n_{i}(x)\}_{i\geq 1}$ that satisfies (\ref{blim}),
we obtain in view of (\ref{rlim})%
\begin{eqnarray}\label{clim}
&&\lim_{n_{i}(x)\rightarrow \infty
}n_{i}(x)^{2}\mathbf{Cov}\{N_{n_{i}(x)}[\varphi
_{1}],N_{n_{i}(x)}[\varphi _{2}]\}  =\mathcal{R}_{V}^{2}(x-\alpha
)\\&& \hskip0.5cm \times \int_{\sigma \times \sigma }\frac{\Delta
\varphi _{1}}{\Delta \lambda }\frac{\Delta \varphi _{2}}{\Delta \lambda }%
\Big( e_{0,0}(\lambda _{1})e_{-1,-1}(\lambda _{2})-e_{0,-1}(\lambda
_{1})e_{0,-1}(\lambda _{2})\Big) d\lambda _{1}d\lambda _{2},
\nonumber
\end{eqnarray}%
where $e_{j,k}(\lambda ),\;j,k\in \mathbb{Z}$ is the density of the
$(jk)$th entry of the resolution of identity
$\{\mathcal{E}_{J_{V}(x)}(d\lambda )\}_{l,m\in \mathbb{Z}}$ of the
limiting Jacobi matrix $J_{V}(x)$, determined by (\ref{JV}).

This result seems rather unusual from the point of view of
traditional probability concepts. Indeed, the covariance of linear
eigenvalue
statistics (\ref{stl}) is of the order $n^{-2}$ rather than of the order $%
n^{-1}$ as in the case of independent identically distributed random
variables, or in a more complex and close to our context case of the
Schrodinger operator with random potential. In the latter case it is
a matter of routine spectral theory argument to show that, say for%
\begin{equation}
\varphi _{z}(\lambda )=(\lambda -z)^{-1},\;\Im z\neq 0,  \label{phiz}
\end{equation}%
where%
\[
N_{n}[\varphi _{z}]=n^{-1}\mathrm{Tr}(H_{n}-z)^{-1}
\]%
and $H_{n}$ is the discrete Schrodinger operator on the interval
$[1,n]$ with a random i.i.d. potential, then%
\[
\lim_{n\rightarrow \infty }n \mathbf{Cov}\{n^{-1}\mathrm{Tr}%
(H_{n}-z_{1})^{-1},n^{-1}\mathrm{Tr}(H_{n}-z_{2})^{-1}\}=C(z_{1},z_{2}),
\]%
where $C(z_{1},z_{2})$ is analytic for $\Im z_{1,2}\neq 0$ and is not
identical zero if the second moment of the potential exists.

On the other hand, it follows from (\ref{clim}) that if $M_n$ is a random $%
n\times n$ random matrix, given by (\ref{M}) -- (\ref{dM}), then
\begin{eqnarray*}
&&\lim_{n_{i}(x)\rightarrow \infty }n_{i}^2(x)\mathbf{Cov}\{n_{i}(x)^{-1}\mathrm{Tr}%
(M_n-z_{1})^{-1},n_{i}(x)^{-1}\mathrm{Tr}(M_n-z_{2})^{-1} \\
&& \hskip1cm =\mathcal{R}_{V}^2(x-\alpha )\left( \frac{\Delta G_{0,0}(x)}{\Delta z}%
\frac{\Delta G_{-1,-1}(x)}{\Delta z}-\left( \frac{\Delta G_{0,-1}(x)}{\Delta
z}\right) ^{2}\right) ,
\end{eqnarray*}%
where%
\[
\frac{\Delta G_{j,k}(x)}{\Delta z}=\frac{1}{z_{1}-z_{2}}\left(
(J_{V(x)}-z_{1})^{-1}-(J_{V(x)}-z_{2})^{-1}\right) _{j,k}.
\]%
Hence the covariance of $N_{n}[\varphi _{z}]=n^{-1}\mathrm{Tr}%
(M_n-z)^{-1}$ is of the order $O(n^{-2})$.

This indicates that the Central Limit Theorem, if any, should be valid not
for $n^{1/2}N_{n}[\varphi _{z}]$ as in the case of i.i.d. random variables,
but for $nN_{n}[\varphi _{z}]$, i.e. for the sum
\[
\mathcal{N}_{n}[\varphi _{z}]=\sum_{l=1}^{n}\varphi (\lambda _{l}^{(n)})
\]%
without a $n$-dependent factor in front.

This was indeed shown in \cite{Jo:98} for the single interval case
$q=1$ and for a rather broad class of test functions. However, as we
have seen above, the case $q=1$ is exceptional, since it is only in
this case asymptotic formulas (\ref{psiV}) and (\ref{rsRSV}) do not
oscillate in $n$ because of the absence of the argument $n\beta $ in
corresponding coefficients of the formulas. Hence, for $q\geq 2$ the
limiting normal law for $\mathcal{N}_{n}[\varphi _{z}]$, if it
exists, could be different for subsequences in (\ref{alim}) having
different limits in $\mathbb{T}^{q-1}$, because its variance depends
on $x\in \mathbb{T}^{q-1}
$:%
\begin{eqnarray*}
&&\lim_{n_{i}(x)\rightarrow \infty
}\mathbf{Var}\{\mathcal{N}_{n_{i}(x)}[\varphi _{z}]\}\\&&
\hskip1cm=\mathcal{R}_{V}^{2}(x-\alpha
)\Big(g_{0,0}(z,x)g_{-1,-1}(z,x)-|g_{0,-1}(z,x)|^{2}\Big),
\end{eqnarray*}%
where%
\[
g_{jk}(z,x)=\int_{\sigma }\frac{(\mathcal{E}_{J_{V}(x)}(d\lambda ))_{j,k}}{%
|\lambda -z|^{2}}.
\]%
However, as is shown in \cite{Pa:05}, the situation with the Central Limit
Theorem for linear eigenvalue statistics of random matrices (\ref{MM}) -- (%
\ref{dM}) is more subtle. Namely, the above scheme of a family of
the Gaussian limiting law with the $x$-dependent variance for
various subsequences in (\ref{blim}) proves to be valid in the
case, where the matrix $J_{V}(x)$ is periodic. In a generic case
of quasiperiodic $J_{V}(x)$ the limiting laws of subsequences
$\{\mathcal{N}_{n_{i}(x)}[\varphi ]\}_{i\geq 1}$ exist but are not
Gaussian.

\appendix

\section{Appendix}
\label{ap}

Here we verify directly that the coefficients of the Jacobi
matrices $J_{\sigma _{g}}(x)$ and $J_{V/g}(x)$ coincide up to a
shift in $x$. We will
consider again the generic case, where the frequencies (\ref{alpha}) and (%
\ref{beta}) are rationally independent, hence $x$ varies over the whole $%
\mathbb{T}^{q-1}$. Besides, we consider only the off-diagonal entries of $%
J_{\sigma _{g}}(x)$ of (\ref{Js}) and $J_{V/g}(x)$ of (\ref{JV}) (note that
the diagonal entries are zero if $V$ is even).

Recall that the coincidence follows also from general results on the
inverse problem of spectral analysis for "finite-band" potentials,
known as the algebro-geometric approach (see e.g.
\cite{Du-Co:90,Le:87,Te:99}). Indeed, since the spectra $J_{\sigma
_{g}}(x)$ and $J_{V/g}(x)$ coincide (see Section 3.2),
$J_{\sigma_g}(x)$ is a isospectral deformation of $J_{V/g}(x)$ and
vice versa, hence, by the inverse problem, one of them can be
obtained by a shift in $x$ of another \cite{Du-Co:90,Le:87,Te:99}).

We begin a direct proof of this assertion by recalling necessary
results of spectral theory of finite band Jacobi matrices and
related facts of complex analysis on Riemann surfaces (see e.g.
\cite{Ap:84,Du-Co:90,Le:87,Te:99}).

Given the set $\sigma $ of (\ref{sigs}), denote $\Gamma $ the two sheeted
(hyperelliptic) Riemann surface, defined by the equation%
\[
w^{2}=R(z),\;R(z)=\prod\limits_{l=1}^{q}(z-a_{l})(z-b_{l}),
\]%
i.e. obtained by pasting together two copies of the complex plane along the
union of the "gaps" $(b_{1},a_{2}),...,(b_{q-1},a_{q}),(b_{q},a_{1})$ of $%
\sigma $, the last gap goes through the infinity point. Let $idp$ be the
normalized differential of the third kind with simple poles of residues $%
\pm 1$, at the infinity points $P_{\pm }$ on each sheet of $\Gamma
$, and
let $U=(U_{1},...,U_{q-1})$ be the vector of $b$-periods of $dp$:%
\begin{equation}
U_{l}=\frac{1}{2\pi }\int_{b_{l}}dp,\;l=1,...,q-1,  \label{Up}
\end{equation}%
where $\{b_{l}\}_{l=1}^{q-1}$ are the so-called $b$-cycles on $\Gamma $.

On the other hand, the integral%
\[
\int_{P_{0}}^{P}idp,\quad P_{0},P\in \Gamma
\]%
with a properly chosen initial point $P_{0}$ can be identified with the
complex Green function $G(z)$ of $\mathbb{C}\setminus \sigma $ with the pole
at infinity (see e.g. \cite{Ap:84}). The real part $g(z)=\Re G(z)$ is
uniquely determined by the requirements to vanish for its limiting values on
$\sigma $ and to be harmonic in $\mathbb{C}\setminus \sigma $ for $g(z)-\log
|z|$. It follows then that if $\nu $ is the unique minimizer of (\ref{Es}),
hence solves the corresponding Euler-Lagrange equation (\ref{Es1}), then%
\[
g(z)=\int_{s}\log |z -\mu |\nu (d\mu )-l_{\sigma }/2.
\]%
This and (\ref{Up}) imply (\ref{alpha}), where $\alpha _{l}-\alpha
_{l+1}=N([a_{l+1},b_{l+1}]),\;l=1,...,q-1$ is the harmonic measure
at infinity of the $(l+1)$th "band" $[a_{l+1},b_{l+1}]$ of $\sigma
$.

Denote $\theta :\mathbb{T}^{q-1}\rightarrow \mathbb{C}$ the Riemann $\theta $%
-function, associated with $\Gamma $. Then according to \cite%
{Wd:69,Ap:84,Pe-Yu:03} the leading coefficient $\gamma _{n}$ of the
polynomial $p_{n}$, where $\{p_{l}\}_{l\geq 0}$ are orthonormal polynomials
on $\sigma $ with respect to weighs, satisfying (\ref{Sz}), is for $%
n\rightarrow \infty $:%
\begin{equation}
\gamma _{n}^{2}=A_{\sigma }e^{nl_{\sigma }}\left[ \frac{\theta (n\alpha
+u(\infty )+d_{\sigma })}{\theta (n\alpha -u(\infty )+d_{\sigma })}+o(1)%
\right] .  \label{gns}
\end{equation}%
Here $l_{\sigma }$ is defined in (\ref{Es1}),
\[
u(z)=\int_{b_{q}}^{z}\omega
\]%
with the integral taken along a path on the first sheet and $\omega
=(\omega _{1},...,\omega _{q-1})$ is the canonical basis of the
differential of the first kind on $\Gamma $,$\ A_{\sigma }$ and
$d_{\sigma }$ do not depend on $n $ but depend on $\sigma ,\;$the
weight, and the points $\zeta _{1},...,\zeta _{q-1}$ of $\Gamma $
that are the poles of the corresponding Baker-Akhiezer function
\cite{Du-Co:90}. In the case, where $\alpha _{l}=m_{l}/q$ with
positive integers $m_{1},...,m_{q}$, hence with a $q$-periodic
$J_{\sigma }(x)$ (see e.g. Section 4), $\zeta _{1}^{\prime
},...,\zeta _{q-1}^{\prime }$ are the eigenvalues of the Dirichlet
problem on the period for the corresponding finite-difference
equation, distributed in a fixed way over the edges of the gaps.
These are in fact the parameters, indexing
representatives of the isospectral family. Another characterization of $%
\zeta _{1}^{\prime },...,\zeta _{q-1}^{\prime }$ is given in \cite{Ap:84},
Theorem W2.

Asymptotic formula (\ref{gns}) and the relation%
\begin{equation}
r_{n}=\gamma _{n}/\gamma _{n+1},  \label{rggs}
\end{equation}%
expressing the off-diagonal entries of a Jacobi matrix via the leading
coefficients of associated orthonormal polynomials, lead to the relation%
\begin{equation}
r_{n}^{2}=e^{-l_{\sigma }}\frac{\theta ((n+1)\alpha -u(\infty )+d_{\sigma
})\theta (n\alpha +u(\infty )+d_{\sigma })}{\theta ((n+1)\alpha +u(\infty
)+d_{\sigma })\theta (n\alpha -u(\infty )+d_{\sigma })}+o(1).  \label{rns}
\end{equation}%
Replacing here $n$ by $n+k$, where $k$ is an arbitrary fixed
integer (in fact $k=o(n)$), and passing to the limit (\ref{alim}),
we obtain for the function $\mathcal{R}_\sigma$ of (\ref{Js}):%
\begin{equation}
\mathcal{R}_{\sigma }(x)=e^{-l_{\sigma }}\frac{\theta (x+\alpha
-u(\infty )+d_{\sigma })\theta (x+u(\infty )+d_{\sigma })}{\theta
(x+\alpha +u(\infty )+d_{\sigma })\theta (x-u(\infty )+d_{\sigma
})}. \label{Rsx}
\end{equation}%
By using the formula%
\begin{equation}
\alpha +2u(\infty )=0,  \label{Rie}
\end{equation}%
that follows from the Riemann bilinear relations
(see e.g \cite{Du-Co:90}, Section 6), we can write%
\begin{equation}
\mathcal{R}_{\sigma }(x)=\mathcal{R}(x+x_{\sigma }),  \label{RRxs}
\end{equation}%
where%
\begin{equation}
\mathcal{R}(x)=e^{-l_{\sigma }}\frac{\theta (x+\alpha )\theta (x-\alpha )}{%
\theta ^{2}(x)}  \label{Rx}
\end{equation}%
and%
\begin{equation}
x_{\sigma}=-u(\infty )+d_{\sigma }.  \label{xp}
\end{equation}%
Consider now the orthonormal polynomials $\{p_{l}^{(n)}\}_{l\geq 0}$ with
respect to varying weights (\ref{wn}) -- (\ref{ortho}). Then we have for the
leading coefficient of $p_{n}^{(n)}$ according to \cite{De-Co:99}, formula (1.63):%
\begin{equation}
\left( \gamma _{n}^{(n)}\right) ^{2}=A_{V}e^{nl_{V}}\left[ \frac{\theta
(n\beta +u(\infty )+d_{V})}{\theta (n\beta -u(\infty )+d_{V})}+o(1)\right]
,\;n\rightarrow \infty ,  \label{gnV}
\end{equation}%
where $l_{V}$ is defined in (\ref{EL1}) -- (\ref{Phi}), $\beta $ is defined
in (\ref{beta}), $u(\infty )$ is the same as in (\ref{gns}), and $A_{V}$ and
$d_{V}$ do not depend on $n$ but depend on $V$ and the points $\zeta
_{1}^{\prime \prime },...,\zeta _{q-1}^{\prime \prime }$, that are zeros of
a certain analytic function on $\mathbb{C}\setminus \sigma $ (see \cite{De-Co:99}%
, formulas (1.26) -- (1.27), (1.30)).

In view of the relations%
\begin{equation}
r_{n+k}^{(n)}=\gamma _{n+k}^{(n)}/\gamma _{n+k+1}^{(n)}  \label{rggV}
\end{equation}%
(cf (\ref{rggs})) and (\ref{rslim}) we need the coefficients $\gamma
_{n+k}^{(n)},\;n\rightarrow \infty ,\;k\in \mathbb{Z}$ fixed (see (\ref{rlim}%
)) in order to find the entries of $J_{V}(x)$. We will find them by
using the same trick as in obtaining (\ref{rlim}). According to the
trick the passage from $n$ to $n+k,\;n\rightarrow \infty ,\;k=o(n)$
can be carried out by passage from the super-index $n$ to $n+k$,
which is equivalent to the infinitesimal change $g\;\rightarrow
\;g+gk/n$ in
the inverse amplitude of the potential. Thus, replacing $V$ by $V/g$ in (\ref%
{gnV}), using the above trick and (\ref{atbd}) -- (\ref{ata}), we obtain (cf
(\ref{gns})):%
\[
\left( \gamma _{n+k}^{(n)}\right)
^{2}=A_{V}e^{nl_{V}+k(gl_{V})^{\prime }} \left[ \frac{\theta
(n\beta +k\alpha +u(\infty )+d_{V})}{\theta (n\beta +k\alpha
-u(\infty )+d_{V})}+o(1)\right] ,\;n\rightarrow \infty ,
\]%
Now, comparing the Euler-Lagrange equations (\ref{EL1}) -- (\ref{Phi}) for (%
\ref{EV}) and (\ref{Es1}) -- (\ref{Es1}) for (\ref{Es}), we find the
relation (cf (\ref{atbd}) -- (\ref{ata})):
\[
l_{\sigma _{g}}=(gl_{V/g})^{\prime }.
\]%
This and (\ref{rggV}) yield for the asymptotics of the off-diagonal entries
of matrix $J^{(n)}$ (\ref{Jn}), associated with $\{p_{l}^{(n)}\}_{l\geq 0}$:%
\begin{eqnarray*}
&&\left( r_{n+k}^{(n)}\right) ^{2} =\left( \gamma
_{n+k}^{(n)}\right)
^{2}\left( \gamma _{n}^{(n)}\right) ^{-2} \\
&&\hskip0.5cm =e^{-l_{\sigma }}\frac{\theta (n\beta +(k+1)\alpha
-u(\infty )+d_{V})\theta (n\beta +k\alpha +u(\infty )+d_{V})}{\theta
(n\beta
+(k+1)\alpha +u(\infty )+d_{V})\theta (n\beta +k\alpha -u(\infty )+d_{V})}%
+o(1).
\end{eqnarray*}%
Passing here to the limit (\ref{blim}), we obtain for the function
$\mathcal{R}_V$ of (\ref{rslim}), determining the off-diagonal
entries of the limiting matrix $J_{V}(x)$:
\[
\mathcal{R}_{V}(k\alpha+x):=\mathcal{R}_{V}(1,l\alpha(1)+x)
=\mathcal{R}(k\alpha + x+x_{V }),
\]
i.e.,
\begin{equation}
\mathcal{R}_{V}(x)=\mathcal{R}(x+x_{V }) \label{RRxV}
\end{equation}%
where $\mathcal{R}$ is defined in (\ref{Rx}), and (cf (\ref{xp}))%
\begin{equation}
x_{V}=-u(\infty )+d_{V/g}.  \label{xpp}
\end{equation}%
Comparing (\ref{RRxs}) and (\ref{RRxV}) we conclude that $\mathcal{R}%
_{\sigma }$ and $\mathcal{R}_{V}$ differ by a shift of argument.

The assertion, formulated at the beginning of the Appendix is proved.

\end{document}